\documentclass{article}
\usepackage{spconf,amsmath,graphicx,hyperref}
\usepackage{xcolor}

\usepackage{cite}
\usepackage{amssymb}
\usepackage{algorithm}
\usepackage{xcolor}
\usepackage{xspace}
\usepackage{booktabs}
\usepackage{url}
\usepackage{pgfplots}
\pgfplotsset{compat=1.18}  
\usepgfplotslibrary{fillbetween}
\usepackage{booktabs}  
\usepackage{algorithm}
\usepackage{algpseudocode}
\usepackage{enumitem}
\usepackage[utf8]{inputenc}
\usepackage[T1]{fontenc}
\usepackage{siunitx}
\newcommand{\best}[1]{\textbf{#1}}      
\newcommand{\up}{$\uparrow$}            
\newcommand{\down}{$\downarrow$}        
\newcommand{\acro}{\textsc{STSM-FiLM}\xspace}

\title{ \acro: A \textsc{FiLM}-Conditioned Neural Architecture \\ for Time-Scale Modification of Speech}
\name{%
\begin{tabular}{c}
\it
Dyah A. M. G. Wisnu$^{\dagger,\ast,\S}$, Ryandhimas E. Zezario$^{\dagger}$, Stefano Rini$^{\ddagger}$, Fo-Rui Li$^{\P,\S}$ \\
Yan-Tsung Peng$^{\ast}$, Hsin-Min Wang$^{\S}$, Yu Tsao$^{\dagger}$
\end{tabular}
}

\address{
    $^{\dagger}$Research Center for Information Technology Innovation, Academia Sinica, Taiwan \\
    $^{\S}$Institute of Information Science, Academia Sinica, Taiwan \\
    $^{\ast}$College of Informatics, National Chengchi University, Taiwan \\
    $^{\ddagger}$Institute of Communications Engineering, National Yang Ming Chiao Tung University, Taiwan \\
    $^{\P}$Dept. of Biomedical Science and Engineering, National Central University, Taoyuan, Taiwan
}
\begin{document}
%
\maketitle
\begin{abstract}
Time-Scale Modification (TSM) of speech aims to alter the playback rate of audio without changing its pitch. 
While classical methods like Waveform Similarity-based Overlap-Add (WSOLA) provide strong baselines, they often introduce artifacts under non-stationary or extreme stretching conditions. 
We propose \acro\xspace-- a fully neural architecture that incorporates Feature-Wise Linear Modulation (FiLM) to condition the model on a continuous speed factor. 
By supervising the network using WSOLA-generated outputs, \acro learns to mimic alignment and synthesis behaviors while benefiting from representations learned through deep learning.
We explore four encoder–decoder variants: STFT-HiFiGAN, WavLM-HiFiGAN, Whisper-HiFiGAN, and EnCodec, and demonstrate that \acro is capable of producing perceptually consistent outputs across a wide range of time-scaling factors. 
Overall, our results demonstrate the potential of FiLM-based conditioning to improve the generalization and flexibility of neural TSM models.
\end{abstract}

\section{Introduction}
\label{sec:intro}
Time-Scale Modification (TSM) is a fundamental audio processing technique that alters the speed of an audio signal without changing its pitch. 
Beyond its use in speech synthesis and audio editing, real-time TSM also has the potential to improve speech intelligibility in noisy environments, offering practical benefits for listeners with hearing difficulties or in challenging acoustic conditions.
Classic TSM algorithms such as the Griffin–Lim algorithm \cite{griffin-lim}, WSOLA (Waveform Similarity Overlap-Add) \cite{WSOLA} and PSOLA (Pitch Synchronous Overlap-Add) \cite{PSOLA} have been widely adopted due to their ability to preserve perceptual quality during stretching or compression.
Generally, these methods rely on handcrafted heuristics, including phase alignment, peak detection, and overlap-add synthesis. 
While effective in many cases, traditional techniques often degrade under non-stationary speech dynamics, suffer from transient smearing, and are sensitive to pitch and formant shifts, especially under extreme time-scale factors \cite{driedger2016review}.

Recently, neural time-scale modification methods \cite{TSM-Net, DiffATSM, FastSpeech2} have emerged as alternatives to classical signal processing. These approaches leverage deep learning to preserve perceptual smoothness, temporal coherence, and intelligibility under varying speed changes. TSM-Net \cite{TSM-Net} uses a convolutional encoder–decoder with a latent Neuralgram representation, allowing interpolation-based time-stretching, while DiffATSM \cite{DiffATSM} applies a diffusion process to synthesize high-quality, scaled speech. Despite these advances, most neural TSM systems remain trained on a limited set of stretch ratios and rarely include explicit conditioning to control the playback rate. Related models such as FastSpeech \cite{FastSpeech2} offer duration control in TTS, but do not address waveform-level TSM.
%

To address this gap, we propose \acro, a fully neural architecture for end-to-end TSM of speech that integrates FiLM-based conditioning into the feature processing pipeline. Instead of directly manipulating audio signals, \acro maps between original and time-scaled audio through deep representations. Specifically, FiLM layers condition the decoder on speed factors, enabling dynamic and continuous control of the playback speed during inference.
Our training target is WSOLA, a strong classical baseline providing consistent supervision. While WSOLA introduces artifacts under extreme factors, it remains effective on clean speech and serves as a practical proxy for training. Our model approximates its behavior while offering a more flexible, fully learnable framework.
We evaluate four encoder–decoder variants—STFT-HiFiGAN, WavLM-HiFiGAN, Whisper-HiFiGAN, and EnCodec—on clean speech. To our knowledge, this is the first work to integrate FiLM into a neural TSM framework.

\section{Proposed Methods}
\label{sec:proposed_method}
\begin{figure}[t]
  \centering
  \includegraphics[width=7.5cm]{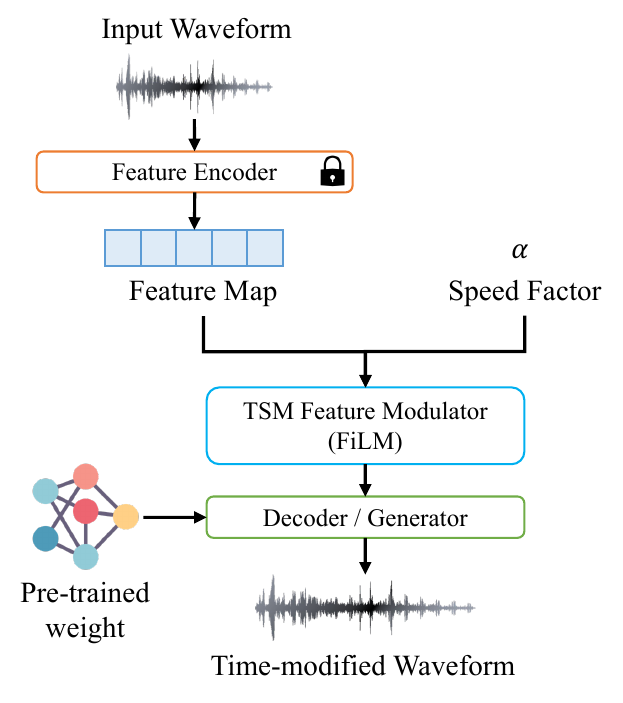}
  \caption{System architecture of the proposed \acro model. The input waveform is encoded into latent feature representations by a feature encoder. The TSM Feature Modulator (FiLM) adapts these features using the speed factor $\alpha$ and pre-trained weights. The decoder/generator then reconstructs the time-scaled waveform from the modulated features.}
  \label{fig:architecture}
\end{figure}

%
%
%
%
The goal of TSM is to alter the duration of an input speech waveform $x(t)$ by a speed factor $\alpha$, without changing its perceived pitch or degrading intelligibility.
Given the input speech signal $x$ and a desired speed factor $\alpha \in (0, \infty)$, the system aims to produce an output $\widehat{x}_\alpha$ such that $\widehat{x}_\alpha$ resembles a naturally spoken version of $x$ at a slower ($\alpha > 1$) or faster ($\alpha < 1$) playback rate.
For this purpose, as shown in Figure \ref{fig:architecture}, we propose \textbf{\acro}, a fully neural TSM model composed of three core components:
\vspace{-4pt}
\begin{itemize}[leftmargin=*,noitemsep]
    \item A \textbf{feature encoder} that transforms raw audio into high-level representations (see Sec. \ref{sec:Feature Encoder}).
    \item A \textbf{FiLM conditioning module} that injects speed factor $\alpha$ as conditioning signals (see Sec. \ref{sec:FiLM Conditioning Module}).
    \item A \textbf{feature decoder} 
    that reconstructs time-scaled waveforms from the modulated features (see Sec. \ref{sec:Feature Decoder}).
\end{itemize}
\vspace{-4pt}
\acro follows an encoder--conditioning--decoder structure. The FiLM-based conditioning module enables the model to dynamically adapt to different time-scale factors at inference time.
The model is trained using WSOLA-generated waveforms as pseudo-ground truth for supervision.

%

\subsection{Feature Encoder}
\label{sec:Feature Encoder}
%
We explore four types of feature encoders with varying levels of abstraction:
\vspace{-4pt}
\begin{enumerate}[leftmargin=*,noitemsep]
    \item \textbf{STFT Encoder}: It computes log-magnitude spectrograms via short-time Fourier transform (STFT), providing a fixed-resolution baseline.
    \item \textbf{WavLM Encoder} \cite{wavlm}: This is a self-supervised transformer-based speech model trained via masked prediction.
    \item \textbf{Whisper Encoder} \cite{whisper}: This is the encoder of the Whisper multilingual ASR and speech translation system.
    \item \textbf{EnCodec Encoder} \cite{encodec}: This is the encoder of a neural codec architecture trained for high-fidelity speech compression. 
\end{enumerate}
\vspace{-4pt}
Each encoder maps the input audio to a sequence of frame-level features $\{f_1, f_2, ..., f_T\}$. For STFT, these are spectrogram bins; for neural encoders, these are learned latent vectors instead.

\subsection{FiLM Conditioning Module}
\label{sec:FiLM Conditioning Module}

To control the time-scaling behavior, we condition the intermediate features on the speed factor $\alpha$ using FiLM layers \cite{film}. A small MLP maps $\alpha$ into affine parameters. 
FiLM learns to scale and shift intermediate features depending on the desired speed, allowing smooth and continuous control of playback speed:
\begin{equation}
(\gamma_\alpha, \beta_\alpha) = \text{MLP}(\alpha).
\end{equation}
These parameters are then used to modulate the features as:
\begin{equation}
\widehat{f}_t = (1 + \gamma_\alpha) \cdot f_t + \beta_\alpha,
\end{equation}
where the offset around 1 helps preserve identity mapping at initialization, improving training stability.

This mechanism allows the model to dynamically scale and shift feature channels in response to the desired speed, allowing for continuous and explicit control during inference. For HiFi-GAN-based variants (with STFT, WavLM, and Whisper features), FiLM is applied at multiple layers within the decoder. For EnCodec, FiLM is injected \textit{between the encoder and the quantizer}, modulating latent codes before quantization.

While FiLM enables adaptive behavior based on $\alpha$, it does not modify the temporal length of the feature sequence. To ensure that the output duration matches the desired speed, we apply linear interpolation along the temporal axis, resizing the feature sequence to the target length. This interpolation step ensures alignment between the conditioned features and the expected duration of the synthesized waveform.

\subsection{Feature Decoder}
\label{sec:Feature Decoder}
The decoder synthesizes the time-scaled waveform $\widehat{x}_\alpha$ from the FiLM-conditioned features:
\vspace{-6pt}
\begin{itemize}[leftmargin=*,noitemsep]
    \item For the STFT, WavLM, and Whisper encoders, we adopt {HiFi-GAN} \cite{hifigan} as the decoder to reconstruct speech from modulated features.
    \item For the EnCodec encoder, we retain the original EnCodec quantizer and decoder to ensure compatibility with learned compression. 
\end{itemize}
\vspace{-6pt}
All decoder variants are trained to minimize reconstruction loss between the generated waveform $\widehat{x}_\alpha$ and the WSOLA-generated pseudo-targets. For HiFi-GAN variants, we also include adversarial losses to enhance perceptual quality\cite{hifigan, wavenet, melgan}.

\subsection{Training and Inference}
Given an input audio $x$ and a speed factor $\alpha$, we generate the WSOLA-based reference $x_\alpha^{\text{WSOLA}}$ and minimize the following loss:
\begin{equation}
\mathcal{L} = \lambda_\text{L1} \cdot \| \widehat{x}_\alpha - x_\alpha^{\text{WSOLA}} \|_1 + \lambda_\text{adv} \cdot \mathcal{L}_\text{GAN} + \lambda_\text{fm} \cdot \mathcal{L}_\text{FM}
\end{equation}
where $\mathcal{L}_\text{L1}$ is the waveform reconstruction loss, $\mathcal{L}_\text{GAN}$ is the adversarial loss from a discriminator, and $\mathcal{L}_\text{FM}$ is the feature matching loss between real and generated speech.
The WSOLA targets act as teacher signals, guiding the model to preserve intelligibility while improving over artifacts introduced by traditional TSM.
%
%

In inference, \acro takes an arbitrary input waveform $x$ and a target speed factor $\alpha$, and produces the time-scaled output $\widehat{x}_\alpha$ without any additional supervision. Thanks to FiLM conditioning, the model supports a continuous range of $\alpha$ values without retraining or reconfiguration.

\section{Experimental Results and Analysis}
\label{sec:experiments}

\noindent{\textbf{Datasets and Setup:}}
We evaluate \acro on two clean speech datasets: \textbf{VCTK} \cite{vctk} (English) and \textbf{TMHINT-QI} \cite{tmhint-qi} (Mandarin). All audio is resampled to 16~kHz. During training, we randomly sample the speed factor $\alpha$ from the range $[0.5, 2.0]$ with a step of 0.1. For each pair $(x, \alpha)$, a WSOLA-processed waveform $x_\alpha^{\text{wsola}}$ is generated as the supervision target.

We train the models for 200k steps using the Adam optimizer with learning rate $8\times10^{-4}$. All models are trained with 32-bit precision.

\vspace{6pt}
\noindent{\textbf{Encoder--Decoder Variants:}}
We benchmark four encoder--decoder configurations under the \acro framework:
\vspace{-6pt}
\begin{itemize}[leftmargin=*,noitemsep]
  \item \textbf{STFT-HiFiGAN}: 1024-dimensional log-mel spectrograms are used as input features, and HiFi-GAN maps the features to a waveform.

  \item \textbf{WavLM-HiFiGAN}: 1024-dimensional features from the 6\textsuperscript{th} layer of WavLM-Large \cite{wavlm} are used as input features, following previous work in voice conversion \cite{VC_1,VC_2}, and HiFi-GAN maps the features to a waveform.

  \item \textbf{Whisper-HiFiGAN}: 1024-dimensional features from the last encoder layer of Whisper Medium \cite{whisper} are used as input features, and HiFi-GAN model maps the features to a waveform.

  \item \textbf{EnCodec}: The official EnCodec encoder--decoder \cite{encodec} is used. FiLM conditioning is applied \textit{between the encoder and quantizer}, and linear interpolation is performed on latent sequences to match the target length $T_\alpha$.
\end{itemize}
\vspace{-3pt}
For all STFT-HiFiGAN, WavLM-HiFiGAN, and Whisper-HiFiGAN, the corresponding HiFi-GAN models are trained on LibriSpeech \cite{librispeech} and COSPRO \cite{cospro}.

\begin{table}[t]
  \centering
  \setlength{\tabcolsep}{5pt}
  \footnotesize
  \caption{Average objective quality and intelligibility assessment across speed factors (0.5–2.0). \up\xspace means higher is better.}
  \label{tab:quality}
  \begin{tabular}{
    l
    S[table-format=1.3] 
    S[table-format=1.3] 
    S[table-format=1.3] 
  }
    \toprule
    \multicolumn{1}{c}{\textbf{System}} &
    \multicolumn{1}{c}{\textbf{PESQ \up}} &
    \multicolumn{1}{c}{\textbf{STOI \up}} &
    \multicolumn{1}{c}{\textbf{DNSMOS \up}} \\
    \midrule
    STFT-HiFiGAN     & \best{2.034} & \best{0.894} & 2.978 \\
    WavLM-HiFiGAN    & 1.924 & 0.891 & \best{2.986} \\
    Whisper-HiFiGAN  & 1.200 & 0.761 & 2.897 \\
    EnCodec          & 1.067 & 0.574 & 2.244 \\
    TSM-Net\cite{TSM-Net}          & 1.417 & 0.741 & 2.287 \\
    \bottomrule
  \end{tabular}
\end{table}

\begin{table}[t]
  \centering
  \setlength{\tabcolsep}{6pt}
  \footnotesize
  \caption{Average objective ASR evaluation across speed factors (0.5–2.0). \down\xspace means lower is better.}
  \label{tab:asr}
  \begin{tabular}{
    l
    S[table-format=1.3] 
    S[table-format=1.3] 
  }
    \toprule
    \multicolumn{1}{c}{\textbf{System}} &
    \multicolumn{1}{c}{\textbf{WER \down}} &
    \multicolumn{1}{c}{\textbf{CER \down}} \\
    \midrule
    STFT-HiFiGAN     & 0.112 & 0.066 \\
    WavLM-HiFiGAN    & \best{0.103} & \best{0.055} \\
    Whisper-HiFiGAN  & 0.198 & 0.332 \\
    EnCodec          & 1.156 & 1.338 \\
    TSM-Net\cite{TSM-Net}          & 0.443 & 0.222 \\
    \bottomrule
  \end{tabular}
\end{table}

\begin{figure}[t]
  \centering
  \includegraphics[width=0.8\columnwidth]{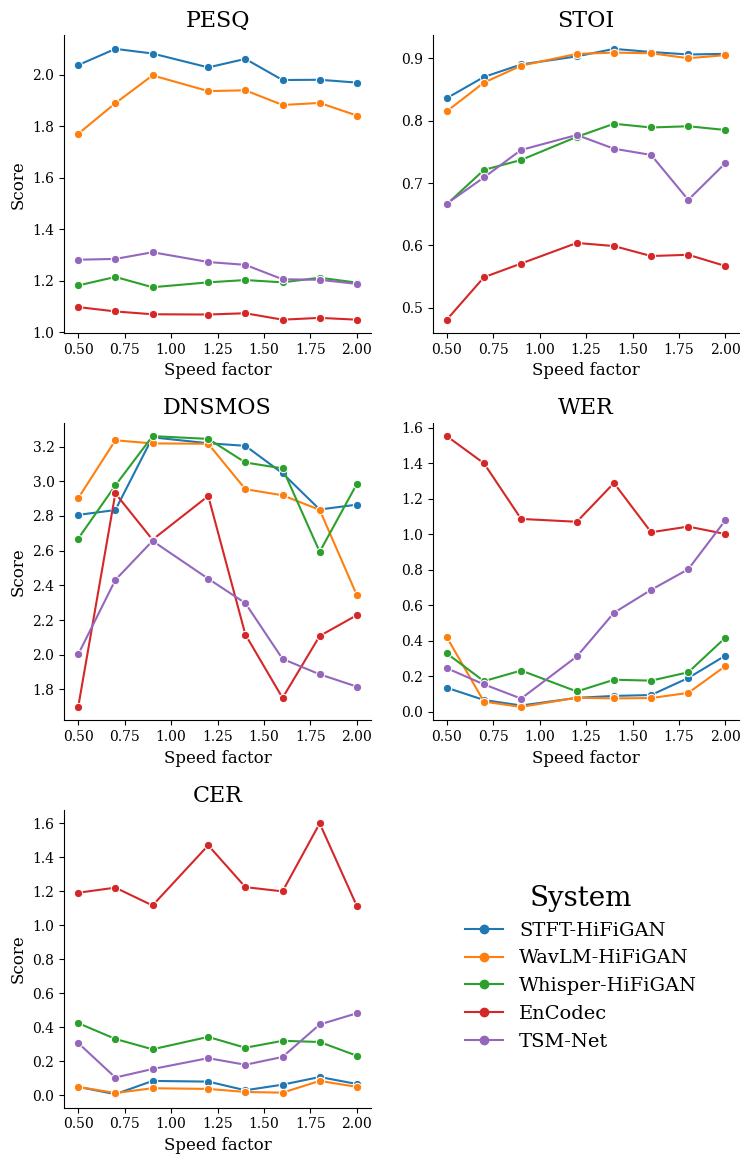}
  \vspace{-0.5cm}
  \caption{Trends of average objective metrics across speed factors (0.5–2.0) for five systems.}
  \label{fig:trend_speed}
\end{figure}

\vspace{6pt}
\noindent{\textbf{Results and Observations:}}
We evaluate the performance of models with PESQ and STOI (quality/intelligibility), DNSMOS (naturalness), and WER/CER from a pretrained ASR for English and Mandarin. Tables \ref{tab:quality}–\ref{tab:asr} report the average performance across different speed factors (0.5–2.0).

STFT-HiFiGAN achieves the best PESQ (2.034) and STOI (0.894), while WavLM-HiFiGAN yields the highest DNSMOS (2.986) and the lowest WER/CER (0.103/0.055). Whisper-HiFiGAN is competitive in DNSMOS (2.897), but suffers in intelligibility and ASR metrics. EnCodec performs worst overall, consistent with degradation from quantization artifacts, while the baseline TSM-Net improves slightly over EnCodec, but remains far behind the HiFiGAN-based models. These results suggest that spectral conditioning (STFT) stabilizes perceptual quality, whereas contextual conditioning (WavLM) captures higher-level phonetic cues, improving naturalness and ASR compatibility. We further hypothesize that the limitations of Whisper-HiFiGAN may arise from padding effects in its conditioning pipeline, whereas those of EnCodec stem from quantization artifacts; a detailed investigation of these mechanisms remains open for future study.



\begin{table}[t]
  \centering
  \caption{Ablation study of FiLM conditioning on WavLM-HiFiGAN at selected speed factors. 
  Reported scores are PESQ / STOI. Higher is better.}
  \label{tab:film_ablation}
  \begin{tabular}{lccc}
    \toprule
    \textbf{WavLM-HiFiGAN} & \textbf{$\alpha=0.7$} & \textbf{$\alpha=1.2$} & \textbf{$\alpha=1.5$} \\
    \midrule
    no FiLM & 1.55 / 0.84 & 1.50 / 0.88 & 1.46 / 0.88 \\
    +FiLM   & \textbf{2.10} / \textbf{0.87} & 
              \textbf{2.03} / \textbf{0.90} & 
              \textbf{2.04} / \textbf{0.90} \\
    \bottomrule
  \end{tabular}
\end{table}

\vspace{6pt}
\noindent{\textbf{FiLM Ablation:}}
Table~\ref{tab:film_ablation} compares WavLM-HiFiGAN with and without FiLM conditioning at selected speed factors. FiLM consistently improves PESQ by roughly 0.5–0.6 points, with the largest gains at the more challenging factors ($\alpha=0.7$ and $\alpha=1.5$). STOI also improves slightly (+0.02–0.03) across all cases. These results confirm that FiLM enhances generalization to different speed factors, particularly by stabilizing perceptual quality under more extreme modifications.

\vspace{6pt}
\noindent{\textbf{Trends across Speeds:}} To further analyze robustness across speed factors, Figure \ref{fig:trend_speed} presents the trends of the objective metrics as a function of the speed factor (0.5–2.0). STFT-HiFiGAN and WavLM-HiFiGAN consistently outperform the other systems, with STFT-HiFiGAN achieving the highest PESQ/STOI and WavLM-HiFiGAN excelling in DNSMOS and ASR metrics. Whisper-HiFiGAN shows moderate DNSMOS but weaker intelligibility, while TSM-Net and EnCodec degrade more noticeably at faster speeds, particularly in WER and CER. These results highlight that HiFiGAN variants, especially STFT- and WavLM-based conditioning, offer greater robustness to temporal distortions compared to baseline approaches.

\vspace{6pt}
\noindent{\textbf{Subjective Listening Test:}}
We conducted a subjective listening test to perceptually assess speech quality. The test used 4 utterances spoken by 4 speakers (2 male, 2 female), modified at six speed factors (0.5, 0.75, 1.25, 1.5, 1.75, 2.0). We compared four systems: TSM-Net, WSOLA, STFT-HiFiGAN, and WavLM-HiFiGAN. Each sample was rated on a five-point mean opinion score (MOS) scale (1 = bad, 5 = excellent). To avoid bias, the four systems were presented in a blind setup so that listeners were unaware of which system generated each sample. Separate tests were conducted for English (LibriSpeech\cite{librispeech}, 15 listeners) and Mandarin (COSPRO\cite{cospro}, 10 listeners). Importantly, all listening materials were drawn from unseen datasets, ensuring that the evaluation reflects out-of-domain generalization.

\begin{table}[t]
  \centering
  \caption{MOS with 95\% confidence intervals 
           for English and Mandarin listening tests (1--5 scale). 
           Higher is better.}
  \label{tab:mos}
  \begin{tabular}{lccc}
    \toprule
    \textbf{System} & \textbf{English} & \textbf{Mandarin} & \textbf{Avg.} \\
    \midrule
    WSOLA         & 4.12 $\pm$ 0.34 & 4.53 $\pm$ 0.34 & 4.33 \\
    TSM-Net       & 1.87 $\pm$ 0.28 & 1.90 $\pm$ 0.55 & 1.89 \\
    STFT-HiFiGAN  & 3.43 $\pm$ 0.22 & 3.70 $\pm$ 0.51 & 3.57 \\
    WavLM-HiFiGAN & \textbf{4.21} $\pm$ 0.30 & \textbf{4.58} $\pm$ 0.24 & \textbf{4.40} \\
    \bottomrule
  \end{tabular}
\end{table}

The results in Table \ref{tab:mos} provide complementary insights to the objective results. While STFT-HiFiGAN achieved the highest PESQ and STOI, listeners consistently rated WavLM-HiFiGAN higher in MOS for both English and Mandarin. This suggests that WavLM features, though less favored by conventional metrics, align more closely with human perception. Notably, even though WavLM-HiFiGAN was trained on WSOLA-processed outputs as pseudo-ground truth, its MOS was comparable to, and in some cases slightly exceeded, that of WSOLA. These findings indicate that neural conditioning can reduce artifacts of classical methods and underscore the value of subjective evaluation for perceptual quality.

\section{Conclusion}
In this paper, we introduce STSM-FILM, a neural framework that uses FiLM conditioning for speech time-scale modification. FiLM allows the model to adapt smoothly across different speed factors and to improve over its WSOLA-processed training targets. In objective tests, STFT-HiFiGAN gave the best signal fidelity, while WavLM-HiFiGAN was stronger in the naturalness and ASR metrics. Subjective listening tests showed that listeners slightly preferred WavLM-HiFiGAN, in some cases even over WSOLA, suggesting that FiLM-based conditioning can capture perceptual qualities not reflected by standard metrics.

\bibliographystyle{IEEEbib}
\bibliography{IEEEabrv,myabbrv,refs} 

\end{document}